\documentstyle[12pt]{article}
\def\fnote#1#2{\begingroup\def\thefootnote{#1}\footnote{#2}
\endgroup}

\begin{document}

\hfill{UTTG-01-00}

\vspace{36pt}

\begin{center}
{\large{\bf {\em A Priori} Probability Distribution of the
Cosmological Constant}}

\vspace{36pt}
Steven Weinberg\fnote{*}{Electronic address:
weinberg@physics.utexas.edu}\\
{\em Theory Group, Department of Physics, University of
Texas\\
Austin, TX, 78712}
\end{center}

\vspace{30pt}

\noindent
{\bf Abstract}

In calculations of the probability distribution for the
cosmological constant, it has been previously assumed that
the {\em a priori} probability distribution is essentially
constant in the very narrow range that is anthropically
allowed.  This assumption has recently been challenged.
Here we identify large classes of theories in which
this assumption is justified.

\vfill

\baselineskip=24pt
\pagebreak
\setcounter{page}{1}

\begin{center}
I. INTRODUCTION
\end{center}

In some theories of inflation$^1$ and of quantum
cosmology$^2$
the observed big bang is just one of an
ensemble of expanding regions in which the cosmological
constant takes various different values.  In such theories
there is a probability distribution for the cosmological
constant: the probability   $d{\cal P}(\rho_V)$ that a
scientific society in any of the expanding regions will
observe a vacuum energy between $\rho_V$ and $\rho_V+\rho_V$
is given by$^{3,4,5}$
\begin{equation}
d {\cal P}(\rho_V)={\cal P}_*(\rho_V){\cal N}(\rho_V)d
4\rho_V\;,
\end{equation}
where ${\cal P}_*(\rho_V)d \rho_V$ is the {\em a priori}
probability that an expanding region will have a vacuum
energy between $\rho_V$ and $\rho_V+d\rho_V$ (to be precise,
weighted with the number of baryons in such regions), and
${\cal N}(\rho_V)$ is proportional to the fraction of
baryons that wind up in galaxies.  (The constant of
proportionality in ${\cal N}(\rho_V)$ is independent of
$\rho_V$, because once a galaxy is formed the subsequent
evolution of its stars, planets, and life is essentially
unaffected by the vacuum energy.)

The  factor ${\cal N}(\rho_V)$
 vanishes except for values of $\rho_V$ that are very small
by the standards of elementary particle physics, because for
$\rho_V$ large and positive there is a repulsive force that
prevents the formation of galaxies$^6$ and hence of stars,
while for $\rho_V$ large and negative the universe
recollapses too fast for galaxies or stars to form.$^7$  The
fraction  of baryons that form galaxies has been
calculated$^5$ for $\rho_V>0$ under reasonable astrophysical
assumptions.    On the other hand, we know little about the
{\em a priori} probability distribution
${\cal P}_*(\rho_V)$.  However, the range of values of
$\rho_V$ in which ${\cal N}(\rho_V)\neq 0$ is so narrow
compared with the scales of energy density typical of
particle physics that it had seemed reasonable in earlier
work $^{4,5}$ to assume that  ${\cal P}_*(\rho_V)$ is
constant within this range, so that  $d {\cal P}(\rho_V)$
can be calculated as proportional to ${\cal N}(\rho_V)d
\rho_V$.  In an interesting recent article,$^8$ Garriga and
Vilenkin have argued that this assumption (which they call
``Weinberg's conjecture'') is generally not valid.  This
raises the problem of characterizing those theories in which
this assumption is valid and those in which it is not.

It is shown in Section II that this assumption is in fact
valid for a broad range of theories, in which the different
regions are characterized by
different values of a scalar field that couples only to
itself and gravitation.  The deciding factor is how we
impose the flatness conditions on the scalar field potential
that are needed to ensure that the vacuum energy is now
nearly time-independent.  If the potential is flat because
the scalar field renormalization constant is very large,
then the {\em a priori} probability distribution of the vacuum
energy is essentially constant within the anthropically
allowed range, for scalar potentials of generic form.  It is
also essentially constant for a large class of other
potentials.
Section III is a digression, showing that the same flatness
conditions ensure tht the vacuum energy has  been roughly
constant since the end of inflation.  Section IV takes up
the sharp peaks in the
{\em
a priori} probability found in
theories of quantum cosmology and eternal inflation.

\vspace{12pt}
\begin{center}
II. SLOWLY ROLLING SCALAR FIELD
\end{center}

One of the possibilities considered by Garriga and Vilenkin
is a vacuum energy that depends on a homogeneous scalar
field $\phi(t)$ whose present value is governed by some
smooth probability distribution.  The vacuum energy is
\begin{equation}
\rho_V=V(\phi)+\frac{1}{2} \dot{\phi}^2\;,
\end{equation}
 and the scalar field time-dependence is given by
\begin{equation}
\ddot{\phi}+3H\dot{\phi}=-V'(\phi)\;,
\end{equation}
where $H(t)$ is the Hubble fractional expansion rate,
$V(\phi)$ is the scalar field potential, dots denote
derivatives with respect to time, and primes denote
derivatives with respect to $\phi$.  Following Garriga and
Vilenkin,$^8$ we assume that at present the scalar field
energy appears like a cosmological constant because the
field $\phi$ is now nearly constant in time, and that this
scalar field energy now dominates the cosmic energy density.
For this to make sense it is necessary for the potential
$V(\phi)$ to satisfy certain flatness conditions.  In the
usual treatment of a slowly rolling
scalar, one neglects the
inertial term $\ddot{\phi}$ in Eq.~(3) as well as the
kinetic energy term $\dot{\phi}^2/2$ in Eq.~(2).  With the
inertial term neglected, the condition that $V(\phi)$ should
change little in a Hubble time $1/H$ is that$^9$
\begin{equation}
V'^2(\phi)\ll 3 H^2 |V(\phi)|\;.
\end{equation}
With the scalar field energy dominating the total cosmic
energy density, the Friedmann equation gives
\begin{equation}
|V(\phi)|\simeq \rho_V \simeq 3H^2/8\pi G\;,
\end{equation}
so  Eq.~(4) requires
\begin{equation}
\left|V'(\phi)\right|\ll \sqrt{8\pi G}\,\rho_V\;.
\end{equation}
(The kinetic energy term $\dot{\phi}^2/2$ in Eq.~(2) can be
neglected under the slightly weaker condition
$$
\left|V'(\phi)\right|\ll \sqrt{18 H^2
\left|V(\phi)\right|}\simeq
\sqrt{48\pi G}\,\rho_V\;,
$$
which is the flatness condition given by Garriga and
Vilenkin.)  There is also a bound on the second derivative
of the potential, needed in order for the inertial term to
be neglected.  With the scalar field energy dominating the
total cosmic
energy density, this condition requires that$^9$
\begin{equation}
\left|V''(\phi)\right|\ll 8\pi G\rho_V\;.
\end{equation}

As Garriga and Vilenkin correctly pointed out, the smallness
of the slope of $V(\phi)$  means that $\phi$ may vary
appreciably even when $\rho_V\simeq V(\phi)$ is restricted
to the very narrow anthropically allowed range of values in
which galaxy formation is
non-negligible.  They concluded that it would be possible
for the {\em a priori} probability ${\cal P}_*(\rho_V)$ to
vary appreciably in this range.  In particular, Garriga and
Vilenkin assumed an {\em a priori} probability distribution
for $\phi$ that is constant in the anthropically allowed
range, in which case the {\em a priori} probability
distribution for
$\rho_V$ is
\begin{equation}
{\cal P}_*(\rho_V)\propto 1/|V'(\phi)|
\end{equation}
which they said could vary appreciably in the anthropically
allowed range.

Though possible, this rapid variation is by no means the
generic case.    As already mentioned, the second as well as
the first derivative of the potential must be small,  so
that  the {\em a
priori} probability density (8) may change little in the
anthropically allowed range.  It all depends on how the
flatness conditions are satisfied.  There are two obvious
ways that one might try to make the potential sufficiently
flat.  Potentials of the first type are of the general form
\begin{equation}
V(\phi)=V_1 f(\lambda\phi)\;,
\end{equation}
where $V_1$ is some large energy density, in the range of
$m_W^{4}$ to $G^{-2}$; the constant $\lambda$ is very small:
and $f(x)$ is some dimensionless function involving no very
large or very small parameters.
Potentials of the second type are of the general form
\begin{equation}
V(\phi)=V_1\left[1-\epsilon\, g(\lambda\phi)\right]\;,
\end{equation}
where $V_1$ is again some large energy density;  $\lambda$
is here some fixed inverse mass, perhaps of order
$\sqrt{G}$; now it is $\epsilon $ instead of $\lambda$ that
is very small; and $g(x)$ is some other dimensionless
function involving no very large or very small parameters.

For potentials (9) of the first type, it is always possible
to meet all observational conditions by taking $\lambda$
sufficiently small, provided that the function $f(x)$ has a
simple zero at a point $x=a$ of order unity, with
derivatives at $a$ of order unity.  Because $V_1$ is so
large, the present value of $\lambda\phi$ must be very close
to the assumed zero $a$ of $f(x)$.  With $f'(a)$ and
$f''(a)$ of order unity, the flatness conditions (6) and (7)
are both satisfied if
\begin{equation}
|\lambda|\ll \left(\frac{\rho_V}{V_1}\right)\sqrt{8\pi G}\;.
\end{equation}
Galaxy formation is only possible for $|V(\phi)|$ less than
an upper bound $V_{\rm max}$ of the order of the mass
density
of the universe at the earliest time of galaxy
formation,$^6$ which in the absence of fine tuning of the
cosmological constant is very
much less than $V_1$.  The
anthropically allowed range of $\phi$ is therefore given by
\begin{equation}
\Delta \phi\equiv |\phi-a/\lambda|_{\rm max}= \frac{V_{\rm
max}}{ |\lambda f'(a)V_1|}\;.
\end{equation}
The fractional change in the {\em a priori} probability
density $1/|V'(\phi)|$ in this range is then
\begin{equation}
\left|\frac{V''(\phi)\Delta\phi}{V'(\phi)}\right|=
\left|\frac{V_{\rm max}}{V_1}\right|\left|\frac{
f''(a)}{f'^2(a)}\right|\;,
\end{equation}
with no dependence on $\lambda$.
As long as the factor $f''(a)/f'^2(a)$ is roughly of order
unity the fractional variation (13) in the {\em a priori}
probability will be very small, as was assumed in references
4 and 5.

This reasoning applies to potentials of the form
$$
V(\phi)=V_1\left[1-(\lambda\phi)^n\right]\;,
$$
which, as already noted by Garriga and Vilenkin, lead to an
{\em priori} probability distribution that is nearly
constant in the anthropically allowed range.  (In this case
$a=1$ and $f''(a)/f'^2(a)=(1-n)/n$.)  But this reasoning
also applies to the ``washboard potential'' that was taken
as a counterexample by Garriga and Vilenkin, which with no
loss of generality can be put in the form:
$$
V(\phi)=V_1\left[1+\alpha\lambda\phi+\beta\sin(\lambda\phi)
\right]\;.
$$
The zero point $a$ is here determined by the condition
$$
1+\alpha a+\beta\sin a=0\;,
$$
and the factor $f''(a)/f'^2(a)$ in Eq.~(13) is
$$
\frac{f''(a)}{f'^2(a)}=\frac{-\beta\sin a}{(\alpha+\beta
\cos a)^2}\;.
$$
If the flatness condition is satisfied by taking $\lambda$
small, with $\alpha$ and $\beta$ of order unity, as is
assumed for potentials of the first kind, then the factor
$f''(a)/f'^2(a)$ in Eq.~(13) is
of order unity unless $\alpha$ and $\beta$ happen to be
chosen so that
$$
\left| 1+\alpha\cos^{-1}\left(\frac{-
\alpha}{\beta}\right)+\beta\sqrt{1-
\frac{\alpha^2}{\beta^2}}\right|\ll 1\;.
$$
Of course it would be possible to impose this condition on
$\alpha$ and $\beta$, but this is the kind of fine-tuning
that would be upset by adding a constant of order $V_1$ to
the potential.  Aside from this exception, for all $\alpha$
and $\beta$ of order unity the factor $f''(a)/f'^2(a)$ is of
order unity, so the washboard potential also yields an {\em
a priori} probability distribution for the vacuum energy
that is flat in the anthropically allowed range.

In contrast, for potentials (10) of the second kind the
flatness conditions are not necessarily satisfied no matter
how small we take $\epsilon$.    Because the present vacuum
energy
is much less than $V_1$, the present value of $\phi$ must be
very close to a value $\phi_\epsilon$, satisfying
\begin{equation}
g(\lambda\phi_\epsilon)=1/\epsilon\;.
\end{equation}
This requires $\lambda\phi_\epsilon$ to be near a
singularity
of the function $g(x)$, perhaps at infinity, so it is not
clear in general that such a potential would have small
derivatives at $\lambda\phi_\epsilon$ for any value of
$\epsilon$.  For instance, for an exponential
$g(x)=\exp(x)$ we have $\phi_\epsilon=- \ln
\epsilon/\lambda$, and
$V'(\phi_\epsilon)$ approaches an $\epsilon$-independent
value proportional to $\lambda$, which is not small unless
we take
$\lambda$ very small, in which case have a potential of the
first kind, for which as we have seen the {\em a priori}
probability density (8) is flat in the anthropically allowed
range.
The flatness conditions {\em are} satisfied for small
$\epsilon$ if $g(x)$ approaches a power $x^n$ for
$x\rightarrow \infty$.  In this case $\phi_\epsilon$ goes as
$\epsilon^{-1/n}$, so $V'(\phi_\epsilon)$ goes as
$\epsilon^{1/n}$ and
$V''(\phi_\epsilon)$ goes as $\epsilon^{2/n}$, both of which
can be made as small as we like by taking $\epsilon$
sufficiently small.

In particular, if the singularity in $g(x)$ at $x\rightarrow
\infty$ consists only of poles in $1/x$ of various orders up
 to $n$
(as is the case for a polynomial of order $n$) then the
anthropically allowed range of $\phi$ is
\begin{equation}
\Big|\phi-\phi_\epsilon\Big|_{\rm max}\approx
\frac{V_m}{V_1\epsilon |g'(\phi_\epsilon)|}\approx
\epsilon^{-1/n}\left(\frac{V_m}{V_1}\right)\;.
\end{equation}
The flatness conditions make this range much greater than
the Planck mass, but the fractional change in the {\em a
priori} probability density (8) in this range is still very
small
\begin{equation}
\left|\frac{V''(\phi_\epsilon)}{V'(\phi_\epsilon)}\right|
\Big|\phi-\phi_\epsilon\Big|_{\rm max}\approx
\frac{V_m}{V_1}\ll 1\;.
\end{equation}
To have a large fractional change in the {\em a priori}
probability distribution in the anthropically allowed range
for potentials of the second type that satisfy the flatness
conditions, we need a function $g(x)$ that goes like a power
as $x\rightarrow \infty$, but has a more complicated
singularity at $x=\infty$ than just poles in $1/x$.
An example is provided by the washboard potential with
$\alpha$ and
$\beta$ very small and $\lambda$ fixed, the case
considered by Garriga and Vilenkin, for which $g(x)$ has an
essential singularity at $x=\infty$.

In summary, the {\em a priori} probability is flat in the
anthropically allowed range for several large classes of
potentials,
while it seems to be not flat only in exceptional cases.

It remains to consider whether the small parameters
$\lambda$
or $\epsilon$ in potentials respectively of the first or
second kind could arise naturally.  Garriga and Vilenkin
argued that a term in a potential of what we have called the
second kind with an over-all factor
$\epsilon\ll 1$ could be naturally produced by instanton
effects.  On the other hand, for potentials of type 1 a
small parameter $\lambda$ could be naturally produced by the
running of a
field-renormalization factor.  The field $\phi$ has a
conventional ``canonical'' normalization, as shown by the
fact that the term $\dot{\phi}^2/2$ in the vacuum energy (2)
and the inertial term $\ddot{\phi}$ in the field equation
(3) have coefficients unity.  Factors dependent on the
ultraviolet  cutoff will therefore be associated with
external $\phi$-lines.  In order for the potential $V(\phi)$
to be expressed in a cut-off independent way in terms of
coupling parameters $g_\mu$ renormalized at a wave-number
scale $\mu$, the field $\phi$ must be accompanied with a
field-renormalization factor $Z_{\mu}^{-1/2}$, which
satisfies a differential equation of the form
\begin{equation}
\mu\frac{d Z_\mu}{d\mu}=\gamma(g_\mu)Z_\mu\;.
\end{equation}
At very large distances, the field $\phi$ will therefore be
accompanied with a factor
\begin{equation}
\lambda=Z^{-1/2}_0=\exp\left\{\frac{1}{2}\int_0^{\mu}
\frac{d\mu'}{\mu'}\gamma(g_{\mu'})\right\}Z^{-1/2}_\mu\;.
\end{equation}
The integral here only has to be reasonably large and
negative  in order for $\lambda$ to be extremely small.

\begin{center}
{\bf III. SLOW ROLLING IN THE EARLY UNIVERSE}
\end{center}

When the cosmic energy density is dominated by the vacuum
energy, the flatness conditions (6) and (7) insure that the
vacuum energy changes little in a Hubble time.  But if the
vacuum energy density is nearly time-independent, then from
the end of inflation until nearly the present it must have
been much smaller than the energy density of matter and
radiation, and under these conditions we are not able to
neglect the inertial term $\ddot{\phi}$ in Eq.~(3).  A
separate argument is needed to show that the vacuum energy
is nearly constant at these early times.  This is important
because, although
there is no observational reason to require $V(\phi)$ to be
constant  at early times, it must have been less than the
energy of radiation at the time of nucleosynthesis in order
not to interfere with the successful prediction of light
element abundances, and therefore at this time must have
been very much less than $V_1$, which we have supposed to be
at least of order $m_W^4$.   For potentials (9) of the first
kind, this means that $\phi$ must have been very close to
its present value at the time of helium synthesis.   Also,
if $\phi$ at the end of
inflation were not the same as $\phi$ at the time of galaxy
formation, then a flat {\em a priori} distribution for the
first would not in general imply a flat {\em a priori}
distribution for the second.

At  times between the end of inflation and the recent past
the expansion rate behaved as  $H=\eta/t$, where $\eta=2/3$
or $\eta =1/2$ during the eras of matter or radiation
dominance, respectively.  During this period, Eq.~(3) takes
the form
\begin{equation}
\ddot{\phi}+\frac{3\eta}{t}\dot{\phi}=-V'(\phi)\;,
\end{equation}
If we tentatively assume that $\phi$ is  nearly constant,
then Eq.~(19) gives for its rate of change
\begin{equation}
\dot{\phi}\simeq-\, \frac{t\,V'(\phi)}{1+3\eta}\;.
\end{equation}
The  change in the vacuum energy from the end of inflation
to the present time $t_0$ is therefore
\begin{equation}
\Delta V\simeq\int_0^{t_0} V'(\phi)\,\dot{\phi}\,dt\simeq -
\,\frac{V'^2(\phi)t_0^2}{2(1+3\eta)}\;.
\end{equation}
The present time is roughly given by $t_0\approx
\eta\sqrt{3/8\pi G\rho_{V0}}$,
so the fractional change in the vacuum energy density since
the end of inflation is
\begin{equation}
\left|\frac{\Delta V}{\rho_{V0}}\right|\approx
 \left(\frac{3\eta^2}{2(1+3\eta)}\right)\,\left(
\frac{V'^2(\phi)}{8\pi G \rho_{V0}^2}\right)\;,
\end{equation}
a subscript zero as usual denoting the present instant.
The factor $3\eta^2/2(1+3\eta)$ is of order unity, so the
inequality (6) tells us that the change in the vacuum
energy during the time since inflation has indeed been much
less
than its present value.

\begin{center}
{\bf III. QUANTUM COSMOLOGY}
\end{center}

In some theories of quantum cosmology the wave function of
the universe is a superposition of terms, corresponding to
universes with different (but time-independent) values for
the vacuum energy $\rho_V$.  It has been argued by Baum$^2$,
Hawking$^2$ and Coleman$^{10}$ that these terms are weighted
with a $\rho_V$-dependent factor, that gives an {\em a
priori} probability distribution with an infinite peak at
$\rho_V=0$, but this claim has been challenged.$^{11}$  As
already acknowledged in references 4 and 5, if this peak at
$\rho_V=0$ is really present, then anthropic considerations
are both inapplicable and unnecessary in solving the problem
of the cosmological constant.

Garriga and Vilenkin$^{8}$ have
proposed a different sort of infinite peak, arising
from a $\rho_V$-dependent rate of nucleation of
sub-universes operating over an infinite time.
Even granting the existence of such a peak, it is not clear
that it really leaves a vanishing normalized probability
distribution at all other values of $\rho_V$.  For instance,
the nucleation rate might depend on the population of 
sub-universes already present, in such a way that the peaks in
the probability distribution are kept to a finite size.
If ${\cal
P}_*(\rho_V)=0$
except at the peak, then anthropic considerations are
irrelevant and the cosmological constant problem is as bad
as ever, since there is no known reason why the peak should
occur in the very narrow range of $\rho_V$ that is
anthropically allowed.  On the other hand, if there is a
smooth background in addition to a peak outside the
anthropically allowed range of $\rho_V$ then the peak is
irrelevant, because no observers would ever
measure such values of $\rho_V$.  In this case the
probability distribution of the cosmological constant can be
calculated using the methods of references 4 and 5.

\begin{center}
{\bf ACKNOWLEDGEMENTS}
\end{center}

I am grateful for a useful correspondence with Alex
Vilenkin.  This research was
supported in part by the
Robert A. Welch
 Foundation and NSF Grant PHY-9511632.

\begin{center}
{\bf REFERENCES}
\end{center}

\begin{enumerate}

\item A. Vilenkin, {\it Phys. Rev.} {\bf D27}, 2848 (1983);
A. D. Linde, {\em Phys. Lett.} {\bf B175}, 395 (1986).

\item E. Baum, {\em Phys. Lett.} {\bf B133}, 185 (1984); S.
W. Hawking, in {\em Shelter Island II - Proceedings of the
1983 Shelter Island Conference on Quantum Field Theory and
the Fundamental Problems of Physics}, ed. R. Jackiw {\em et
al.} (MIT Press, Cambridge, 1995); {\em Phys. Lett.} {\bf
B134}, 403 (1984); S. Coleman, {\it Nucl. Phys.} {\bf B
307}, 867 (1988).

\item An equation of this type was given by A. Vilenkin,
{\em Phys. Rev. Lett.} {\bf 74}, 846 (1995); and in {\em
Cosmological Constant and the Evolution of the Universe}, K.
Sato, {\em et al.}, ed. (Universal Academy Press, Tokyo,
1996) (gr-qc/9512031), but it was not used in a calculation
of the mean value or probability distribution of $\rho_V$.

\item S. Weinberg, in {\em Critical Dialogs in Cosmology},
ed. by N. Turok (World Scientific, Singapore, 1997).

\item H. Martel, P. Shapiro, and S. Weinberg, {\em Ap. J.}
{\bf 492}, 29 (1998).

\item S. Weinberg, {\em Phys. Rev. Lett.} {\bf 59}, 2607
(1987).

\item J. D. Barrow and F. J. Tipler, {\it The Anthropic
Cosmological Principle} (Clarendon Press, Oxford, 1986).

\item J. Garriga and A. Vilenkin, Tufts University preprint
astro-ph/9908115, to be published.

\item P. J. Steinhardt and M. S. Turner, {\it Phys. Rev.}
{\bf D29}, 2162 (1984).

\item S. Coleman, {\em Nucl. Phys.} {\bf B 310}, 643 (1988).

\item W. Fischler, I. Klebanov, J. Polchinski, and L.
Susskind, {\em Nucl. Phys.} {\bf B237}, 157 (1989).

\end{enumerate}

\end{document}